\documentclass[11pt,a4paper]{article}

\usepackage[margin=1in]{geometry}
\usepackage{amsmath,amssymb,amsthm}
\usepackage{graphicx}
\usepackage{hyperref}
\usepackage{braket}
\usepackage[T1]{fontenc}
\usepackage{lmodern}
\usepackage[backend=biber,style=phys,biblabel=brackets]{biblatex}
\usepackage{authblk} 
\usepackage{orcidlink}

\begin{document}
	
	\title{Regularization of Vortex Core Size in Photon BECs due to Harmonic Trap: An Analytical Approach}
	
	\author[1,*]{Joshua Krauß\,\orcidlink{0009-0003-1297-9551}}
	\author[1,2]{Tabitha Bates\,\orcidlink{0009-0006-7726-0435}\thanks{Present address: School of Physics and Astronomy, University of Birmingham, Birmingham B15 2TT, United Kingdom}}
	\author[3]{Francisco Ednilson Alves dos Santos}
	\author[1]{Axel Pelster\,\orcidlink{0000-0002-5215-0348}}

	\affil[1]{\small Department of Physics and Research Center OPTIMAS, University Kaiserslautern-Landau, Erwin-Schrödinger Straße 46, 67663 Kaiserslautern, Germany}
	\affil[2]{\small School of Physics and Astronomy, University of St.~Andrews, St.~Andrews KY16 9SS, United Kingdom}
	\affil[3]{\small Departamento de Física, Universidade Federal de S\~{a}o Carlos, S\~{a}o Carlos, S\~{a}o Paulo, 13565-905, Brazil}
	\affil[*]{\small Author to whom any correspondence should be addressed. E-mail: \href{mailto:jkrauss@rptu.de}{jkrauss@rptu.de}, \href{mailto:txb624@student.bham.ac.uk}{txb624@student.bham.ac.uk}, \href{mailto:santos@ufscar.com}{santos@ufscar.com}, \href{mailto:axel.pelster@rptu.de}{axel.pelster@rptu.de}}
	\date{\vspace{-5ex}}
	
	\maketitle
	
	\begin{abstract}
		Quantized vortices are a hallmark of superfluidity. However, in a photon Bose--Einstein condensate, the photon-photon interaction is so weak that in a homogeneous system the healing length exceeds the experimentally achievable size of the condensate itself. Here we show that a harmonic confinement regularizes the vortex size to experimentally achievable scales. Moreover, such an external confinement closely resembles the standard experimental setup of a pumped dye-filled microcavity. We model the condensate via a complex Gross--Pitaevskii equation and obtain an approximate dynamical single-vortex solution
		by applying the recently proposed projection optimization method. The latter generalizes the variational approach of closed systems to open-dissipative systems without assuming a specific form for the condensate phase. The radial photon flow, which is characteristic for driven-dissipative systems, yields a definition of the vortex core size based on the competition of gain and loss. However, this condition reproduces the heuristic closed system definition now based on physical grounds. A subsequent linear stability analysis shows that interaction, as well as pumping and loss can drive the system to an unstable regime. In this way one can fundamentally distinguish between closed and open-dissipative systems.
	\end{abstract}
	
	\vspace{1em}
	
	\section{Introduction}\label{sec:introduction}
	Bosonic quantum gases can undergo a process in which the ground state is macroscopically occupied. This phenomenon is known as Bose--Einstein condensation and has been observed experimentally in various systems such as, for instance, in atomic or molecular quantum gases~\cite{Cornell_Science1995,Ketterle_PRL1995,Greiner_Nature2003}, magnon gases~\cite{Hillebrands_Nature2006}, plasmonic lattices~\cite{Torma_Nature2018}, exciton polaritons~\cite{Kasprzak_Nature2006} or photonic systems~\cite{Klaers_Nature2010,Nyman_Nature2024,Piezcarka_Nature2024}. The latter systems have attracted particular interest because of their open-dissipative nature. 
	Experimentally, a Bose--Einstein condensate (BEC) of pure light was realized in a Vertical-Cavity Surface-Emitting Laser~\cite{Nyman_Nature2024,Piezcarka_Nature2024,Fainstein_NatPhot2024} and earlier in a dye-filled microcavity~\cite{Klaers_Nature2010}. In this work, we will focus on the latter realization. This setup consists of a curved-mirror microcavity, which provides a ground state and imposes a harmonic confinement for the effectively two-dimensional photons. In addition, the cavity is filled with a dye, usually Rhodamine 6G, which thermalizes the particles of light through repeated emission and absorption cycles. The dye is pumped by an external laser field, providing a dynamical equilibrium and therefore a quasi-conserved particle number. One major advantage is that the underlying microscopic physics is well understood~\cite{Kirton_PRA2015,Kirton_PRL2013,Radonjic_NJP2018}.\\
	Whether a condensate of light is a superfluid is still an open question. Among the hallmarks of superfluidity, quantized vortices are perhaps the most direct ones. These topological defects are characterized by a quantized circulation around a phase singularity, where the condensate density has to vanish~\cite{Pethick_Smith_2008}. In a homogeneous conservative condensate, the distance over which the density recovers from the defect is set by the healing length and, thus, by the strength of the particle-particle interaction~\cite{Pethick_Smith_2008}. For photon BECs this interaction arises from a weak Kerr nonlinearity, which turns out to be so small~\cite{Dung_NatPhot2017,Klaers_APB2011} that the healing length would exceed the size of any realistic condensate. Thus, a vortex would simply not fit in the system.\\
	However, the open-dissipative nature of a photon condensate changes this picture qualitatively. Numerical studies of nonequilibrium condensates have shown that the interplay of pumping and losses can change the size of the vortex core, and that these vortices differ structurally from their closed-system counterparts~\cite{Keeling_PRL2008,Gladilin_NJP2017,Gladilin_PRL2020,Krauss_PRR2025}. In particular, in addition to the tangential velocity field known from closed systems, a radial component emerges, so that the flow of particles around the vortex core becomes spiral rather than circular. The vortex then acts as a source of particles~\cite{Gladilin_NJP2017}. This spiral structure was recently analytically described for a homogeneous photon condensate~\cite{Krauss_PRR2025}, where the core size was found to follow a coherence length 
	that depends on both the pumping and the losses of the particles. Consequently, in driven-dissipative condensates, the core size depends not only on the interaction strength. That length nevertheless maintains its interaction dependence, so that the core size remains macroscopically large. So far a crucial experimental detail has not been taken into account, namely the external harmonic trapping potential imposed by the cavity mirrors. Such a trap introduces a second length scale, which is independent of the interaction strength. Therefore, it is natural to ask whether it regularizes the vortex core to an experimentally accessible size.\\
	Answering this requires an analytical treatment that complements numerical studies~\cite{Keeling_PRL2008,Gladilin_NJP2017,Gladilin_PRA2021,Gladilin_PRL2020} and exposes how the vortex structure depends on the system parameters. For closed condensates the analytical description is based on the variational optimization method~\cite{Lewenstein_PRL1996,Lewenstein_PRA1997}. With a suitably chosen trial wave function, a minimization of the underlying action determines the variational parameters. In this way the low-lying collective modes of a condensate~\cite{Lewenstein_PRL1996,Lewenstein_PRA1997,Fetter_PRM2009} and the structure and dynamics of vortices~\cite{Chevy_PRL2002,Cidrim_PRA2018} have been obtained. For open-dissipative condensates, however, this procedure is not directly applicable. Neither energy nor particle number is conserved and, therefore, an action from which the equations of motion could be derived requires careful usage of the Schwinger-Keldysh formalism~\cite{Sieberer_RMP2025}. One way to overcome this obstacle is the cumulant optimization method, where the equations of motion for the cumulants of the condensate wave function are derived and evaluated self-consistently~\cite{Stein_NJP2019}. However, this method is limited to wave functions that are described by a finite number of cumulants, which excludes vortices. A different method is given by the projection optimization method~\cite{Krauss_PRR2025}, which we will use in this work. It is based on projecting the underlying equation of motion onto a manifold spanned by trial parameters introduced in a wave function ansatz. This method does not require the underlying action and generalizes both methods mentioned above.\\
	Applying this method to a harmonically trapped photon condensate including a singly charged vortex described by the complex Gross--Pitaevskii equation~\cite{Keeling_PRL2008}, we find that the confinement indeed regularizes the vortex core. Its size is determined by the oscillator length and is therefore within experimental reach. Moreover, the radial velocity component provides an intrinsic definition of the vortex core size via the distance at which the outward current induced by the vortex reverses into the inward flow from the harmonic confinement. This physical mechanism reproduces the heuristical criterion introduced in a closed-system condensate~\cite{Cidrim_PRA2018}. Finally, a linear stability analysis reveals that interaction, as well as pumping and loss can destabilize the stationary state through an oscillatory instability of the breathing mode. Comparing with the vortex-free scenario shows the same behavior, so we conclude that the vortex instability is inherited from the condensate itself. In summary, we show that a vortex in a photon condensate is both structurally well defined and experimentally observable. This provides an important step towards understanding quantized vortices, and thereby superfluidity, in open-dissipative quantum fluids.\\
	We structured this work as follows. In Sec.~\ref{sec:model} we introduce the complex Gross--Pitaevskii equation describing a harmonically trapped photon condensate. Section~\ref{sec:method} summarizes the projection optimization method and specifies the trial ansatz. Section~\ref{sec:Single_vortex} contains our projection optimization results for the singly charged vortex. We discuss the stationary state including the vortex core size and a linear stability analysis. The corresponding vortex-free solution, including its breathing mode, is derived in App.~\ref{app:no_vortex}, while App.~\ref{app:hydrodynamics} provides a hydrodynamic interpretation for the radial flow.

	\section{Model}\label{sec:model}
	In this section, we introduce the model employed to describe a photon BEC. Experimentally, the condensation of photons occurs within a curved mirror cavity, where the ground-state energy of the photons can be tuned by the mirror geometry \cite{Klaers_Nature2010}. Additionally, in the paraxial approximation, the cavity effectively reduces the dimensionality of the system from three to two and the photons are mapped onto massive bosons trapped in a harmonic potential \cite{Klaers_Nature2010,Klaers_APB2011}.\\
	A key prerequisite for achieving condensation is the presence of a dye medium, typically Rhodamine 6G. Through repeated cycles of absorption and emission processes, the photons thermalize to the temperature of the dye molecules. Hence, the thermalization of the photons is inherited from the medium \cite{Ricard_CPL1972,Shank_RMP1975}, which is governed by the Kennard-Stepanov relation \cite{Kennard,Stepanov}. This relation connects the emission coefficient $B_{21}$ to the absorption coefficient $B_{12}$ of the dye via the Boltzmann factor
	\begin{equation}
		\label{eq:KS_relation}
		B_{12}(\omega) = B_{21}(\omega) e^{\hbar\beta(\omega-\omega_{\rm ZPL})}\,.
	\end{equation}
	Here, $\beta = 1/k_{\rm B} T$ denotes the inverse temperature, with $T=300\,{\rm K}$ representing room temperature, and $\omega_{\rm ZPL}=550\,\text{THz}$ stands for the zero-phonon line frequency of Rhodamine 6G. Additionally, an external laser field is applied to pump photons into the cavity, thereby compensating for mirror-induced photon losses.\\
	One of the simplest models for this system relies on rate equations \cite{Krauss_EPJST2026}, which can be derived from microscopic principles \cite{Kirton_PRL2013,Kirton_PRA2015,Radonjic_NJP2018,Ozturk_PRA2019,Bode_PRR2024} and accurately capture both the photon density and the molecular dynamics throughout the condensation process until a steady state is reached. However, this rate-equation description lacks information about the phase of the condensate, hence, it cannot describe collective modes or topological defects such as vortices. To this end, we consider the photon wave function $\Psi$, which follows the complex Gross--Pitaevskii equation (cGPE)~\cite{Keeling_PRL2008, Krauss_PRR2025}, and given by
	\begin{equation}
		\label{eq:cGPE}
		i\hbar\partial_t\Psi({\bf r},t) = \left\{-\frac{\hbar^2}{2m}\boldsymbol{\nabla}^2+\frac{1}{2}m\omega^2 {\bf r}^2+g|\Psi({\rm r},t)|^2+\frac{i}{2}\left[\gamma-\Gamma\left|\Psi({\bf r},t)\right|^2\right]\right\}\Psi({\bf r},t)\,.
	\end{equation}
	In this framework, the photons are trapped in an external harmonic potential with frequency $\omega$, while $m$ and $g$ denote the effective photon mass and the interaction strength, respectively. Typical values are $\omega = 2\pi\cdot 40\,\text{GHz}$ \cite{Klaers_Nature2010} and $m=10^{-5}\cdot m_{\rm e}$ \cite{Bloch_NatRev2022} with the electron mass $m_{\rm e}$, which are both set by the mirror geometry \cite{Klaers_Nature2010}, as well as 
	$g=10^{-5}\cdot \hbar^2/m$ \cite{Dung_NatPhot2017}.
	In contrast to a conservative BEC, the additional imaginary term in Eq.~\eqref{eq:cGPE} accounts for the non-equilibrium nature of the system. It represents the balance between the effective pumping strength $\gamma$ and density-dependent losses with the rate $\Gamma$. Notably, while Eq.~\eqref{eq:cGPE} was originally utilized for exciton-polariton condensates \cite{Keeling_PRL2008} and can be derived within the Schwinger-Keldysh framework \cite{Sieberer_RMP2025}, it also follows heuristically for photon BECs from extending the rate-equation description \cite{Krauss_PRR2025}. Furthermore, this heuristic derivation allows to map the microscopic parameters of the rate equations \cite{Krauss_EPJST2026} onto the loss rate in Eq.~\eqref{eq:cGPE}, yielding a typical value of $\Gamma = 10^{-4}\cdot \hbar^2/m$. In the following analysis, we will treat the pumping strength $\gamma$ as a dynamically tunable parameter, closely mirroring standard experimental procedures.

	\section{Optimization Method and Ansatz}\label{sec:method}
	While the previous section introduced the equation of motion for the photon field, the resulting cGPE (\ref{eq:cGPE}) lacks an exact analytical solution. Consequently, it is typically addressed through numerical simulations \cite{Keeling_PRL2008,Gladilin_NJP2017,Gladilin_PRL2020,Gladilin_PRA2021} or approximate analytical methods \cite{Stein_NJP2019}. In this work, we focus on the latter.\\
	In closed systems, approximate analytical solutions are commonly found by minimizing the action using Hamilton's principle \cite{Lewenstein_PRL1996,Lewenstein_PRA1997}. However, for the open-dissipative system considered here, defining an action is non-trivial due to the lack of particle and energy conservation. Therefore, we employ the projection optimization method proposed in Ref.~\cite{Krauss_PRR2025} in order to find approximate analytical solutions to Eq.~\eqref{eq:cGPE}. This method generalizes the standard optimization method for closed systems \cite{Lewenstein_PRL1996,Lewenstein_PRA1997} to their open-dissipative counterpart. Moreover, unlike the cumulant optimization method~\cite{Stein_NJP2019}, it is not restricted to vortex-free systems and does not rely on the knowledge of an action. Instead, it projects the equation of motion (EOM), which is given here by the cGPE~\eqref{eq:cGPE}, onto a manifold spanned by the optimization functions $\boldsymbol{\lambda}({\bf r},t) = (\lambda_1({\bf r},t),\lambda_2({\bf r},t),\ldots)$. These functions are introduced via the trial ansatz for the wave function $\Psi({\bf r},t) \approx \psi({\bf r},t,\boldsymbol{\lambda}({\bf r},t))$. Two key points warrant emphasis here. First, if the action is known, the projection optimization method is mathematically equivalent to the Hamilton optimization method. Secondly, extending the framework of Ref.~\cite{Krauss_PRR2025}, we use here spatiotemporally dependent trial functions. This generalization allows to determine the full dynamics of the ansatz without making {\it a priori} assumptions about its temporal behavior.\\
	The resulting projection optimization equations (POEs), that determine these optimization functions, are given by
	\begin{equation}
		\label{eq:POE}
		\Braket{\text{EOM}\left[\psi,\psi^*\right]|\frac{\delta \psi(\textbf{r},t,\boldsymbol{\lambda}({\bf r},t))}{\delta\lambda_i({\bf r^\prime},t^\prime)}} + \text{c.c.}= 0\,.
	\end{equation}
	Here, $\Braket{\bullet|\bullet}$ denotes a general projection onto the manifold, which must be suitably defined for the specific system under consideration. In this work, we define the projection via the standard scalar product
	\begin{equation}
		\label{eq:scalar_product}
		\braket{f|g} = \int_0^\infty dt\int_{\mathbb{R}^2} d{\bf r}\, f^*({\bf r},t)\cdot g({\bf r},t)\,.
	\end{equation}
	In order to apply the projection optimization method to our system, we make the following trial ansatz for the two-dimensional photon wave function~\cite{Cidrim_PRA2018,Ostrovskaya_PRA2012}
	\begin{equation}
		\label{eq:ansatz}
		\psi(r,\phi,t) = \sqrt{\frac{N(t)}{\pi|\ell|!q(t)^{2|\ell|+1}}}\,r^{|\ell|}e^{-\frac{r^2}{q(t)^2}+i\left[\ell\phi + \varphi_{\rm R}(r,t)\right]}\,e^{-\frac{i}{\hbar}\mu(t)}\,.
	\end{equation}
	Here $\ell\in\mathbb{Z}$ denotes an integer that defines phase quantization, arising from the single-valuedness of the wave function \cite{Pethick_Smith_2008}. This allows us to describe both the vortex-free case $\ell = 0$ and a single vortex with vortex charge $\ell \neq 0$ at the trap center. Note that Eq.~\eqref{eq:ansatz} does not make any assumption on the functional form of the unknown phase $\varphi_{\rm R}$, which generalizes both the ansätze used for a closed~\cite{Lewenstein_PRL1996,Lewenstein_PRA1997,Chevy_PRL2002,Fetter_PRM2009} and an open-dissipative system~\cite{Stein_NJP2019,Ostrovskaya_PRA2012}. Furthermore, we include time-dependent optimization functions for the photon number $N(t)$, the length scale $q(t)$, the generalized chemical potential $\mu(t)$, and the radial phase $\varphi_R(r,t)$. Note that the length scale $q(t)$ represents the width of the condensate in the absence of a vortex, while in the case of a single vortex it determines the spatial shape of the vortex. Moreover, we use polar coordinates $r,\phi$ to introduce the velocity field $\boldsymbol{v}(r,\phi,t)$, which is related to the phase of the wave function by $\boldsymbol{v}(r,\phi,t) = \hbar \boldsymbol{\nabla}[\ell\phi + \varphi_{\rm R}(r,t)]/m$. This yields, according
	to the Helmholtz theorem \cite{Helmholtz}, two distinct contributions to the velocity field. The tangential component $\boldsymbol{v}_{\rm T}(\phi) = \hbar\ell\boldsymbol{e}_\phi/(mr)$ describes the quantized rotation of the condensate around the trap center, and the radial component $\boldsymbol{v}_{\rm R}(r,t) = \hbar v_{\rm R}(r,t) \boldsymbol{e}_r/m$ describes a radial flow of photons with a yet-to-be-determined magnitude $v_{\rm R}(r,t)$. Although the tangential part is assumed to remain unaffected by the open-dissipative nature of the system \cite{Pethick_Smith_2008,Krauss_PRR2025}, the existence of a radial velocity component is essential for stabilizing the condensate, as it governs the redistribution of particles \cite{Gladilin_NJP2017,Stein_NJP2019}. Note that the radial velocity component is used in closed systems only to describe the dynamics of collective modes~\cite{Lewenstein_PRL1996,Lewenstein_PRA1997} and vanishes in a stationary state, whereas in an open-dissipative system this component is necessary due to the constant flux of particles in the system even in a steady state. \\
	It is important to emphasize that the projection optimization method allows for determining this radial photon flow without requiring {\it a priori} assumptions or a specific ansatz for the radial velocity field. In the following, we demonstrate that the knowledge of the condensate's density profile alone is sufficient to determine the corresponding radial velocity component.

	\section{Singly-Charged Vortex Solution}\label{sec:Single_vortex}
	In this section, we apply the projection optimization method to a system with a singly-charged vortex located at the trap center in order to investigate the influence of the open-dissipative parameters on the vortex's core size and stability. The section is organized as follows. In Sec.~\ref{sec:POE_vortex} we derive 
	the yet unknown functions of the trial ansatz~\eqref{eq:ansatz} for the specific case of $\ell=\pm 1$
	from the POEs. Next, we analyze in Sec.~\ref{sec:SS_vortex} the stationary state of this vortex solution, which allows us to determine the vortex core size and examine how it depends on the system's inherent open-dissipative nature. Finally, in Sec.~\ref{sec:Stability_vortex} we investigate the stability of this solution, and demonstrate that tuning the pump and loss rates as well as the interaction strength can induce an instability of the singly-charged vortex. Additionally, in both sections 
	\ref{sec:SS_vortex} and
	\ref{sec:Stability_vortex}
	we compare the vortex solution to the vortex-free counterpart, determined and discussed in App.~\ref{app:POE_no_vortex}, and show the differences in both configurations.
	\subsection{Vortex Projection Optimization Equations}\label{sec:POE_vortex}
	To fully characterize the dynamics of the singly-charged vortex, we evaluate the POEs for the length scale $q(t)$, the particle number $N(t)$, the chemical potential $\mu(t)$, and the radial phase $\varphi_{\rm R}(r,t)$. Notably, the first three of these equations explicitly depend on the spatial profile of the radial phase. Therefore, we first determine the POE for this function. Applying the standard functional derivative rules alongside the defining relation $v_{\rm R} = \hbar \partial_r \varphi_{\rm R} / m$ introduced in Sec.~\ref{sec:model}, the POE reduces to the following ordinary differential equation
	%.	
	\begin{equation}
		\label{eq:vel_ode}
		\partial_r v_{\rm R} + \frac{3q^2 - 2r^2}{r q^2}\,v_{\rm R} = \frac{\gamma}{\hbar} - \frac{\Gamma N}{\pi\hbar} \frac{r^2}{q^4}\, e^{-\frac{r^2}{q^2}} - \frac{\partial_t N}{N} - \frac{2r^2 - 4q^2}{q^2}\,\frac{\partial_t q}{q}\,.
	\end{equation}
	Formally solving for the spatial profile of the velocity field yields 
	\begin{equation}
		\label{eq:vel_vortex0}
		\begin{aligned}
			v_{\rm R} = \frac{r\partial_t q}{q} &+ \left(
			\frac{\gamma}{\hbar} - \frac{\partial_t N}{N} - \frac{\Gamma N}{4\pi\hbar q^2}\right) \frac{q^4}{2r^3}\,e^{\frac{r^2}{q^2}}\\
			&+\frac{q^2}{2r^3}\left[\frac{\Gamma N}{4\pi\hbar q^2}\,e^{-\frac{r^2}{q^2}}\left(q^2 + 2r^2 + \frac{2r^4}{q^2}\right) -  
			\left(\frac{\gamma}{\hbar}  - \frac{\partial_t N}{N}\right)
			\left(q^2+r^2\right)\right]\,.
		\end{aligned}
	\end{equation}
	Note that integrating Eq.~\eqref{eq:vel_ode} with respect to the radial coordinates led to a spatial integration constant, which was determined by imposing a Dirichlet boundary condition at the origin, i.e.~demanding that the radial velocity field must vanish at this point~\cite{Krauss_PRR2025,Gladilin_NJP2017,Gladilin_PRL2020}. This spatial integration constant is given by
	\begin{equation}
		\label{eq:int_const_vortex}
		C_v = \frac{q^4}{2}    
		\left(
		\frac{\gamma}{\hbar} -\frac{\partial_t N}{N} - \frac{\Gamma N}{4 \pi\hbar q^2}
		\right)\,,
	\end{equation}
	and could still be a function of time.
	With this integration constant the radial velocity~\eqref{eq:vel_vortex0} vanishes, indeed, at the origin, but still has an exponential divergence for large distances. However, using Eq.~\eqref{eq:vel_vortex0}, we can now determine the POE for the chemical potential, which yields a differential equation for the particle number
	\begin{equation}
		\label{eq:N_vortex}
		\partial_t N = N\left(\frac{\gamma}{\hbar} - \frac{\Gamma N}{4\pi\hbar q^2}\right)\,.
	\end{equation}
	We recognize that Eq.~\eqref{eq:N_vortex}
	has the consequence that 
	the spatial integration constant \eqref{eq:int_const_vortex} vanishes, i.e.~we have $C_v = 0$. Additionally, with this we obtain that 
	the radial velocity field~\eqref{eq:vel_vortex0} 
	no longer diverges exponentially in the far field. Thus, the particle number equation \eqref{eq:N_vortex}  
	renormalizes the radial velocity field so that all unphysical divergences cancel, and with \eqref{eq:N_vortex} the radial velocity~\eqref{eq:vel_vortex0} simplifies to
	\begin{equation}
		\label{eq:vel_vortex}
		v_{\rm R} = \frac{r\partial_t q}{q} + 
		\frac{\Gamma N}{8 \pi \hbar r^3}
		\left[e^{-\frac{r^2}{q^2}}\left(q^2 + 2r^2 + \frac{2r^4}{q^2}\right) - \left(q^2 + r^2\right)\right]\,.
	\end{equation} 
	We complete the description of the radial flow by determining the corresponding phase. To this end, we spatially integrate Eq.~\eqref{eq:vel_vortex}, once again imposing a Dirichlet boundary condition at the origin such that the radial phase vanishes. This results in
	\begin{equation}
		\label{eq:phase_vortex}
		\varphi_{\rm R} = \frac{m}{\hbar}\left\{\frac{r^2 \partial_t q}{2 q} + 
		\frac{\Gamma N}{16 \pi \hbar}
		\left[1 - \gamma_{\rm EM} - \ln\left(\frac{r^2}{q^2}\right) + \frac{q^2}{r^2} - {\rm E}_1\left(\frac{r^2}{q^2}\right) - e^{-\frac{r^2}{q^2}}\left(2 + \frac{q^2}{r^2}\right)\right] \right\}\, ,
	\end{equation}
	where $\gamma_{\rm EM}= 0.57721\dots$ denotes the Euler-Mascheroni constant and ${\rm E}_1$ stands for the exponential integral. Having determined the radial velocity and phase, we can now evaluate the remaining POEs, which read for the width
	\begin{equation}
		\label{eq:q_vortex}
		\partial^2_t q = \frac{\hbar^2}{m^2 q^3} - \omega^2 q + \frac{g N}{8\pi m q^3} + \frac{\Gamma \partial_t N}{32 \pi \hbar q} + 
		\frac{\left[27\ln(4/3) - 11\right]\Gamma^2N^2}{3456 \pi^2\hbar^2q^3}\,,
	\end{equation}
	and the chemical potential
	\begin{equation}
		\label{eq:mu_vortex}
		\partial_t\mu = m\omega q\partial^2_t q + \frac{\hbar^2\omega}{m q^2} + m\omega ^3 q^2 + \frac{\omega gN}{4\pi q^2} - \frac{\left[2\ln(2) - 1\right]
			m\omega \Gamma \partial_t N}{32\pi\hbar} + \frac{\left[27\ln(4/3) - 4\right]m\omega \Gamma^2 N^2}{3456\pi^2\hbar^2q^2}\,,\,\,\,\,\,
	\end{equation}
	respectively.
	It is worth noting that the POEs for the photon number and the chemical potential yield equations for the respective opposite optimization functions. We also remark that the chemical potential is completely determined by Eq.~\eqref{eq:mu_vortex}, depending on the solutions of Eqs.~\eqref{eq:N_vortex} and \eqref{eq:q_vortex}. Consequently, our system of coupled differential equations effectively reduces to two coupled equations for $N$ and $q$, which we will analyze in the following.

	\subsection{Steady-State Vortex Solution}\label{sec:SS_vortex}
	One inherent property of weakly-interacting driven-dissipative systems, compared to their closed counterpart, is the existence of a stationary state. This state can often be tuned by the parameters of pumping and dissipation. In the case of the single vortex considered here, the steady state provides information about the anatomy of the vortex, in particular how large it is.\\
	As the structure of the vortex depends on the solution of the POEs~\eqref{eq:N_vortex},~\eqref{eq:q_vortex}, and~\eqref{eq:mu_vortex}, we start our analysis with determining the stationary solution of these equations. To this end, we make the steady-state ansatz of a time-independent particle number and length scale, i.e.~$N(t)=N_0$, $q(t)=q_0$, and a linear time dependence for the chemical potential $\mu(t)=\omega\mu_0 t$. With this, the optimization functions from Sec.~\ref{sec:POE_vortex} become time-independent optimization parameters. Moreover, the POEs of these parameters become algebraic equations and can be solved analytically. The respective solutions read
	\begin{align}
		\label{eq:q_vortex_ss}
		q^2_{0,\pm} &= \frac{-\frac{g\gamma}{2\hbar\omega\Gamma} \pm \sqrt{\left(\frac{g\gamma}{2\hbar\omega\Gamma}\right)^2 - \frac{\gamma^2}{2\hbar^2\omega^2}\left[\ln\left(\frac{4}{3}\right) - \frac{11}{27} - \frac{8\hbar^2\omega^2}{\gamma^2}\right]}}{\frac{\gamma^2}{4\hbar^2\omega^2}\left[\ln\left(\frac{4}{3}\right) - \frac{11}{27} - \frac{8\hbar^2\omega^2}{\gamma^2}\right]}\,\ell_\omega^2\,,\\
		\label{eq:N_vortex_ss}
		N_0 &= \frac{4\pi\gamma q^2_{0,\pm}}{\Gamma}\,,\\
		\label{eq:mu_vortex_ss}
		\mu_0 &= \frac{\hbar^2}{m q^2_{0,\pm}} + m\omega^2q^2_{0,\pm} + \frac{g\gamma}{\Gamma} + \frac{m\gamma^2 q^2_{0,\pm}}{8\hbar^2}\left[\ln\left(\frac{4}{3}\right) - \frac{4}{27}\right]\,,
	\end{align}
	where $\ell_\omega=\sqrt{\frac{\hbar}{m\omega}}$ denotes the oscillator length, a typical length scale in a harmonically trapped system. We note that Eq.~\eqref{eq:N_vortex_ss} represents the global particle balance, i.e.~the total gain compensates the total loss of particles.
	\begin{figure}[t!]
		\centering
		\includegraphics[width=\columnwidth]{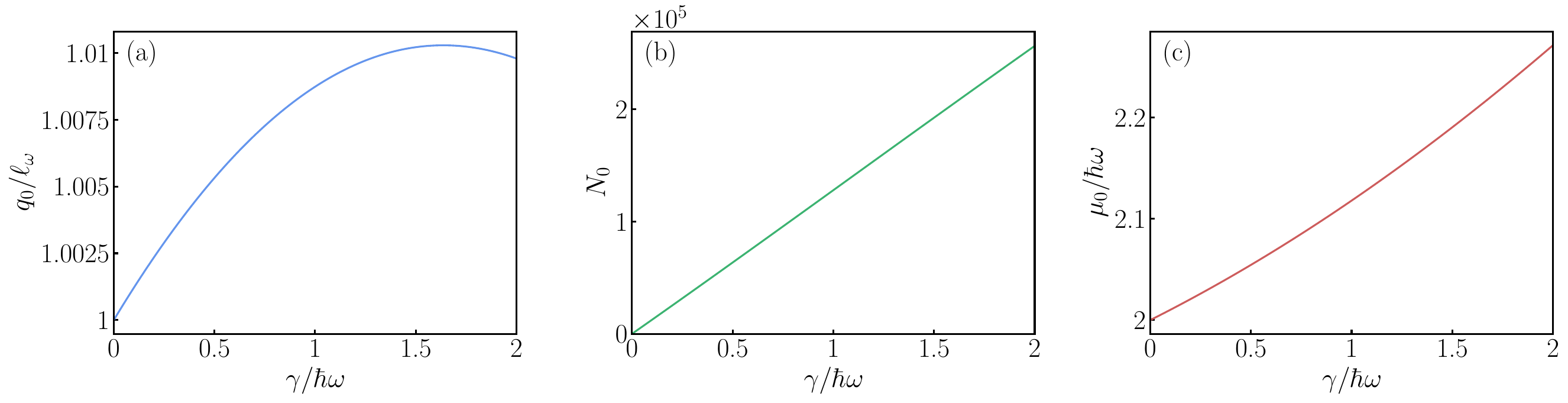}
		\caption{Projection optimization parameters from Eqs.~\eqref{eq:q_vortex_ss}--\eqref{eq:mu_vortex_ss} for steady-state of vortex solution as function of pumping $\gamma$ for $\Gamma = 10^{-4}\cdot \hbar^2/m$ and $g = 10^{-5}\cdot \hbar^2/m$. Panel (a) shows  length scale $q_0$, panel (b) photon number $N_0$, and panel (c) chemical potential $\mu_0$. }
		\label{fig:proj_params_ss}
	\end{figure}
	First of all, we note that there are two solutions for the length scale~\eqref{eq:q_vortex_ss} as indicated by the different signs of the square root.
	Expanding with respect to small pumping strength allows 
	distinguishing between the mathematical and the physical solution. From the limit
	\begin{equation}
		\label{eq:q_SS_limit}
		q_{0,\pm}^2 \xrightarrow{\gamma\rightarrow 0} \left(\mp 1 + \frac{g\gamma}{4\hbar\omega\Gamma} \right)\ell_\omega^2
	\end{equation}
	we read off that the plus solution becomes negative for vanishing pumping and, therefore, corresponds to an unphysical complex valued length, whereas the minus solution always remains positive for small pumping strengths and, thus, corresponds to a real-valued length scale. With this we conclude that the latter denotes the physical solution $q_{0,-}\equiv q_0$, which will be  considered later.\\
	Figure~\ref{fig:proj_params_ss}(a) shows that the physical solution of the length scale increases with increasing pumping strength until a maximum is reached, and after that the length decreases again. This nonmonotic behavior reveals that the validity range of our ansatz~\eqref{eq:ansatz} is limited. Namely, according to Fig.~\ref{fig:proj_params_ss}(b) the photon number linearly increases, as expected from the rate-equation description in the deeply condensed regime \cite{Krauss_EPJST2026}. Therefore, considering both Fig.~\ref{fig:proj_params_ss}(a) and (b) together suggests that with increasing particle number the total size of the system decreases. As we do not expect such a behavior the validity range of our ansatz~\eqref{eq:ansatz} is restricted to small pumping strengths $\gamma \lesssim \hbar\omega$. However, this is also expected from Ref.~\cite{Keeling_PRL2008}. There, it was shown numerically that for fixed losses an increase of the pumping rate ultimately deforms the density profile due to the emergence of particle currents. Then the density can no longer be approximated by a Gaussian profile in this limit. Hence, as the ansatz~\eqref{eq:ansatz} does not accommodate such a deformation, it is restricted to the regime in which these currents do not yet modify the density distribution. Finally, we notice that the chemical potential shown in Fig.~\ref{fig:proj_params_ss}(c) also increases for increasing pumping strengths as expected. However, we note that the chemical potential converges for vanishing pumping to the value $\mu_0 = 2\hbar\omega$. This can be explained by taking the zero-pumping limit of Eqs.~\eqref{eq:mu_vortex_ss} and~\eqref{eq:mu_no_vortex_ss}, yielding
	\begin{equation}
		\label{eq:chem_pot_expand}
		\mu_0 \xrightarrow{\gamma\rightarrow 0} \left(1 + \left|\ell\right|\right)\hbar\omega\,,
	\end{equation}
	and also coinciding with the closed system expectation.
	Therefore, in the case of a singly-charged vortex it is increased compared to the case without a vortex.\\
	Now, we investigate the spatial profile of the vortex. To this end, we start our analysis with the radial velocity and the phase profile, which are in the steady state known from Eqs.~\eqref{eq:vel_vortex} and~\eqref{eq:phase_vortex} by
	\begin{align}
		\label{eq:vel_vortex_ss}
		v_{\rm R,0} &= \frac{\gamma q_0^2}{2\hbar r^3}\left[e^{-\frac{r^2}{q_0^2}}\left(q_0^2 + 2r^2 + \frac{2r^4}{q_0^2}\right) - \left(q_0^2 + r^2\right)\right]\,,\\
		\label{eq:phase_vortex_ss}
		\varphi_{\rm R,0} &= \frac{\gamma m q_0^2}{4\hbar^2}\left[1 - \gamma_{\rm EM} - \ln\left(\frac{r^2}{q_0^2}\right) + \frac{q_0^2}{r^2} - {\rm E}_1\left(\frac{r^2}{q_0^2}\right) - e^{-\frac{r^2}{q_0^2}}\left(2 + \frac{q_0^2}{r^2}\right)\right]\,.
	\end{align}
	First of
	\begin{figure}[t!]
		\centering
		\includegraphics[width=\columnwidth]{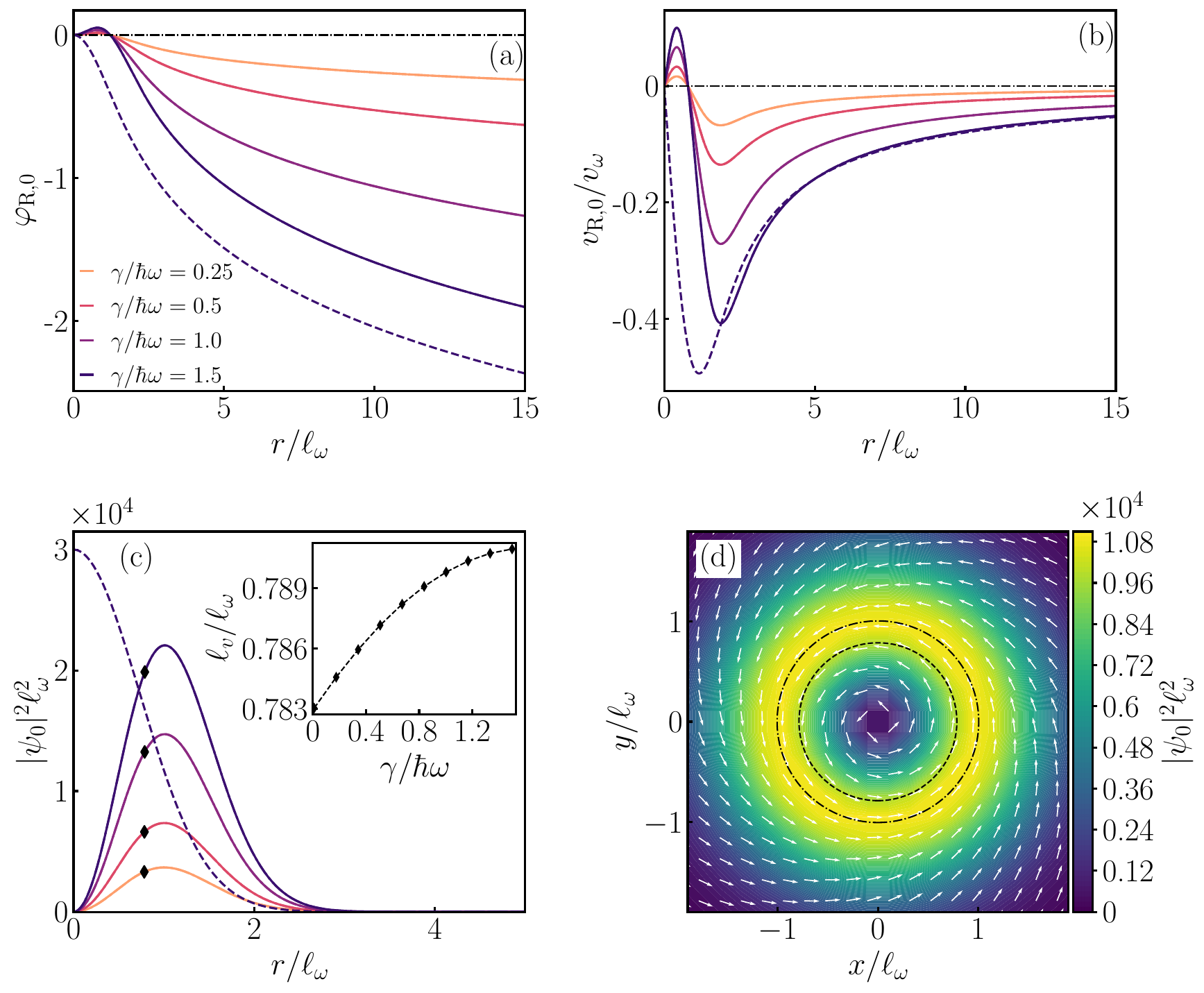}
		\caption{Panel (a) and (b) show steady-state solution of radial phase and velocity profile according to Eqs.~\eqref{eq:vel_vortex_ss} and~\eqref{eq:phase_vortex_ss} for different pumping values as well as $\Gamma = 10^{-4}\cdot \hbar^2/m$ and $g = 10^{-5}\cdot \hbar^2/m$. In Panel (b) radial velocity profile is given in units of oscillator velocity $v_\omega = \sqrt{\frac{\hbar\omega}{m}}$. In both panels solid lines depict vortex solution and dashed line one representative solution of vortex-free case determined in App.~\ref{app:no_vortex_ss}. Dash-dotted horizontal line represents zero-value for better visualization of positive and negative parts. Panel (c) follows the same convention as Panels (a) and (b),  depicting density distribution according to Eq.~\eqref{eq:ansatz} including Eqs.~\eqref{eq:q_vortex_ss}--\eqref{eq:mu_vortex_ss}. Black points indicate occurrence of zero radial velocity and inset shows these points as function of respective pumping strength. Panel (d) illustrates two-dimensional density distribution. White arrows show normalized velocity field, where dashed (dash-dotted) circle denotes  points of zero radial velocity (maximum density). }
		\label{fig:vortex_anatomy}
	\end{figure}
	all, we notice that both expressions have a positive and a negative contribution, the positive (negative) one dominating at small (large) distances. 
	The results \eqref{eq:vel_vortex_ss} and~\eqref{eq:phase_vortex_ss} are shown in Fig.~\ref{fig:vortex_anatomy}(a) and (b)  for different pumping strengths. Close to the trap center the radial velocity is positive, so that photons are transported away from the vortex core, a behavior already known from the trap-free case~\cite{Krauss_PRR2025}. However, beyond a well-defined distance the current reverses and photons flow inwards.\\
	Expanding Eq.~\eqref{eq:vel_vortex_ss} and its vortex-free counterpart~\eqref{eq:vel_no_vortex_ss} around the trap center, yields
	\begin{align}
		\label{eq:near_axis_vortex}
		v_{\rm R,0} &\xrightarrow{r\rightarrow 0} \frac{\gamma r}{4\hbar}\,,\\
		\label{eq:near_axis_no_vortex}
		v_{\rm R,0}^{\rm (vf)} &\xrightarrow{r\rightarrow 0} -\frac{\gamma r}{2\hbar}\,,
	\end{align}
	where the superscript ${\rm (vf)}$ refers to the vortex-free case. We note that both limits are independent of $q_0$, $\Gamma$ and $g$. Therefore, the phase singularity reverses the sign of the near-origin flow from a linearly compressing flow towards the trap center to a linearly expanding one, at exactly half the rate. Equations~\eqref{eq:near_axis_vortex} and~\eqref{eq:near_axis_no_vortex} represent the initial slopes of the corresponding curves in Fig.~\ref{fig:vortex_anatomy}(b), and also agree with the near-origin behavior predicted for exciton-polariton condensates~\cite{Ostrovskaya_PRA2012}. The radial velocity field is determined by the POE~\eqref{eq:vel_ode}, which coincides with the continuity equation~\eqref{eq:continuity_eq}, derived in App.~\ref{app:hydrodynamics}. Therefore, in the case of a known density distribution the radial flow of particles is purely determined by the competition of pumping and losses. In the steady state, gain and loss balance locally at the density $\gamma/\Gamma$. The phase singularity decreases the density below this value at the trap center, so that the vortex acts as a net source, whereas the surrounding dense ring acts as a net sink. Photons pumped in the vicinity of the vortex core therefore flow outwards, whereas photons created in the dilute tail flow inwards. Both currents are then absorbed in between, so that particles accumulate in a ring around the core. The radial velocity vanishes, where gain and loss of particles cancel. The cancellation, however, does not appear at the maximum density; instead, it is achieved before that. A detailed explanation of this behavior follows from the hydrodynamic description derived in App.~\ref{app:hydrodynamics}.\\
	Determining where the radial velocity~\eqref{eq:vel_vortex_ss} becomes zero across the physical pumping interval shows that the density at this point is always $90.3\%$ of its maximum value, where we used again the same values for $g$ and $\Gamma$ as previously. Note that due to the non-trivial radial dependence of Eq.~\eqref{eq:vel_vortex_ss} this evaluation can only be done numerically. Moreover, as the chosen ansatz~\eqref{eq:ansatz} does not include a separate length scale that determines the size of the vortex, we define its core size as the distance where gain compensates for loss, so where the radial velocity~\eqref{eq:vel_vortex_ss} becomes zero. This allows us to determine that the size of the vortex, which we denote with $\ell_v$, is of the order of the oscillator length, i.e. $\ell_v \approx \ell_\omega$. For the experimental parameters introduced in Sec.~\ref{sec:model} the oscillator length is given by $\ell_\omega \approx 7.5\,\mu\text{m}$, hence, the vortex core size is $\ell_v \approx 5.8\,\mu\text{m}$ and lies well inside the experimental optical resolution. Furthermore, we note that the closed-system counterpart of the ansatz~\eqref{eq:ansatz} faces the same problem of not having a length scale that determines the vortex core size. Therefore, it was heuristically defined as the distance from the core, where the density reaches $90\%$ of its maximum value~\cite{Cidrim_PRA2018}, which coincides with the definition chosen above. The only difference in our case is that the vortex core size is based on physical grounds and can be determined once the radial velocity profile is known.

	\subsection{Stability of Vortex Solution}\label{sec:Stability_vortex}
\begin{figure}[t!]
		\centering
		\includegraphics[width=\columnwidth]{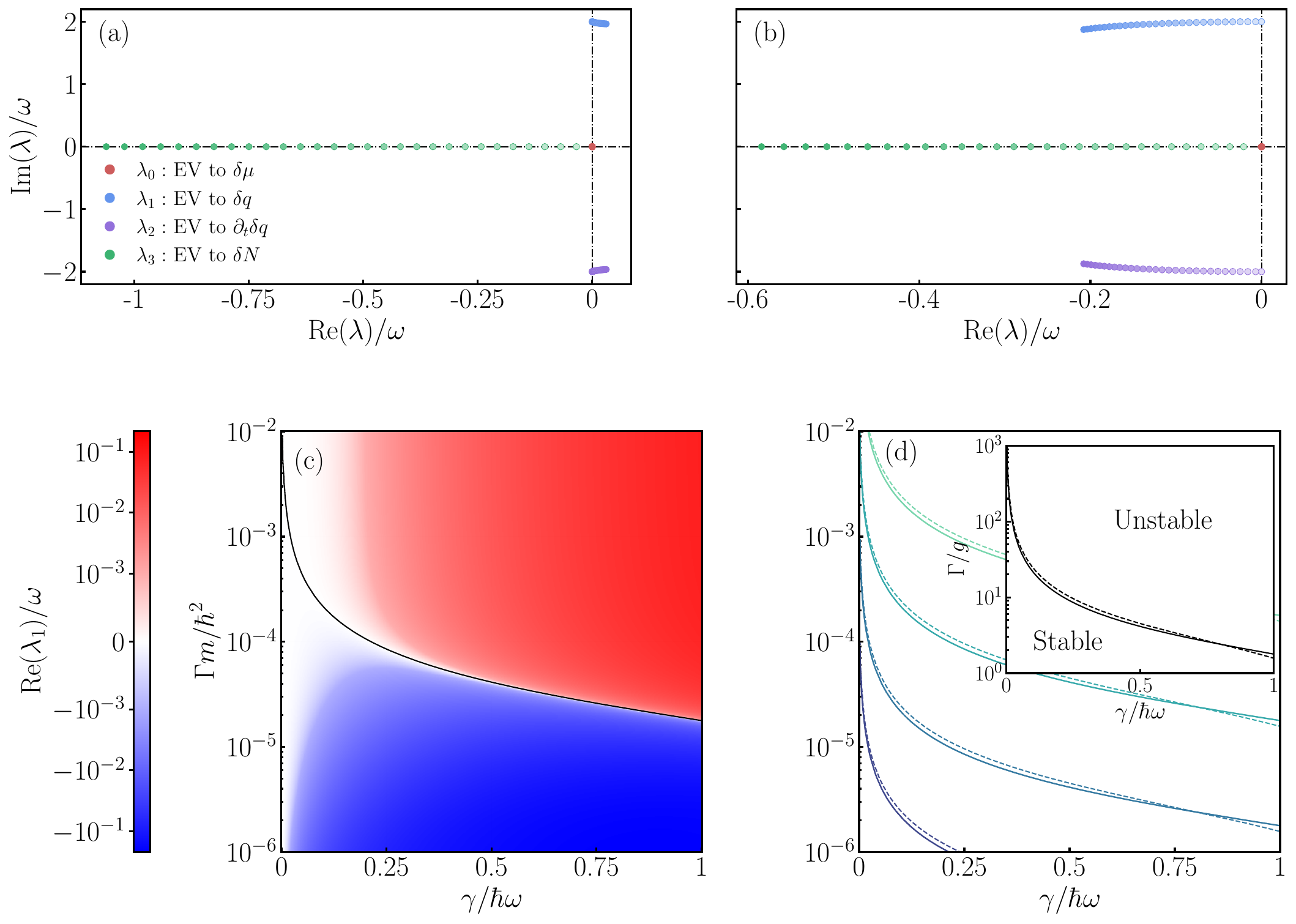}
		\caption{Panel (a) and (b) show eigenvalues of matrix $A$ as function of pumping strength. Bright colors denote small pumping with smallest value $\gamma/\hbar\omega = 0$, whereas strong colors denote large pumping with maximum being $\gamma/\hbar\omega = 1$. In both Panels we have $g = 10^{-5}\cdot \hbar^2/m$, whereas in left (right) Panel $\Gamma = 10^{-3}\cdot \hbar^2/m$ ($\Gamma = 10^{-7}\cdot \hbar^2/m$) is used. Panel (c) depicts  real part of eigenvalue $\lambda_1$ in semi-logarithmic scaling for $g = 10^{-5}\cdot \hbar^2/m$. Black solid line indicates points of vanishing real part. Panel (d) illustrates line of zero real part of eigenvalue $\lambda_1$ for different interaction strengths; from top to bottom interaction strength values are $g = 10^{-4}\cdot \hbar^2/m$, $10^{-5}\cdot \hbar^2/m$, $10^{-6}\cdot \hbar^2/m$, $10^{-7}\cdot \hbar^2/m$. Solid lines correspond to vortex, whereas dashed lines illustrate vortex-free case, see App.~\ref{app:no_vortex}. Inset shows collapse of lines from main figure to single line.}
		\label{fig:vortex_stability}
	\end{figure}

	In the previous section, we determined the steady-state solution of the POEs~\eqref{eq:N_vortex}, \eqref{eq:q_vortex} and~\eqref{eq:mu_vortex} and showed that the ansatz~\eqref{eq:ansatz} for fixed $g=10^{-5}\cdot \hbar^2/m$ and $\Gamma=10^{-4}\cdot \hbar^2/m$ faces restrictions in its validity due to the absence of possible density deformations. Now, in this section, we look at the stability of the steady-state solution in the valid pumping interval found in Sec.~\ref{sec:SS_vortex} and show whether any further restriction induced by the open-dissipative nature is relevant. To this end, we consider small perturbations around the steady state by considering expansions for projection optimization functions $N(t) = N_0 + \delta N(t)$, $q(t) = q_0 + \delta q(t)$, and $\mu(t) = \omega\mu_0 t + \delta\mu(t)$. Inserting this expansion into  POEs~\eqref{eq:N_vortex}, \eqref{eq:q_vortex} and~\eqref{eq:mu_vortex}, and expanding the resulting equations up to linear order in the perturbations, leads to a first-order differential equation, which determines the perturbations, of the form $\partial_t \boldsymbol{x} = A \boldsymbol{x}$ with $\boldsymbol{x}^{\rm T} = \left(\delta N(t), \delta q(t), \partial_t \delta q(t), \delta \mu(t)\right)$ and a time-independent matrix $A\in {\rm Mat}(4\times 4, \mathbb{R})$. The formal solution of this differential equation is given by
	\begin{equation}
		\label{eq:stability_solution}
		\boldsymbol{x}(t) = e^{A t}\boldsymbol{x}_0\,,
	\end{equation}
	where $\boldsymbol{x}_0$ denotes some initial condition and $e^{A t}$ stands for the matrix exponential of the matrix $A$. Note that, due to the huge size of the matrix, we do not show it explicitly. The solution of Eq.~\eqref{eq:stability_solution} and especially the stability of the steady-state solution from Sec.~\ref{sec:SS_vortex} is solely determined by the eigenvalues of the matrix $A$, which are in general complex-valued. Therefore, we achieve a stable stationary state only if every real part of the eigenvalues either vanishes or is negative. The eigenvalues $\lambda_i$ with $i=1,2,3,4$ are determined by the roots of the characteristic polynomial $\chi_A$, which is given by
	\begin{equation}
		\label{eq:characteristic_polynomial}
		\chi_A(\lambda) = \omega^4 \frac{\lambda}{\omega}\left[\left(\frac{\lambda}{\omega}\right)^3 + \frac{\gamma}{\hbar\omega}\left(\frac{\lambda}{\omega}\right)^2 + \left(4-\frac{\gamma^2}{4\hbar^2\omega^2}\right)\frac{\lambda}{\omega} + \frac{\gamma^2 g\ell_\omega^2}{\hbar^2\omega^2\Gamma q_0^2}\left(1 + \frac{4\hbar\omega\Gamma\ell_\omega^2}{g\gamma q_0^2}\right)\right]\,.
	\end{equation}
	We immediately see that one eigenvalue is always zero, i.e.~$\lambda_0 = 0$, which corresponds to the eigenvalue induced by the chemical potential. As the chemical potential decouples from the other projection optimization functions, see Eq.~\eqref{eq:mu_vortex}, this also holds for the respective perturbation yielding a vanishing eigenvalue. Note that this reflects the global $U(1)$ symmetry of the cGPE~\eqref{eq:cGPE}. The remaining eigenvalues for the perturbations of the length scale and the photon number are now computed numerically and shown in Fig.~\ref{fig:vortex_stability}(a) and (b) together with the eigenvalue stemming from the chemical potential for different pumping strengths and two different loss parameters. First we notice that the eigenvalue corresponding to the particle number is always real-valued, negative, and decreases for increasing pumping strength in both loss rate cases. Especially, we see that in the limit of vanishing pumping this eigenvalue becomes zero, which is equivalent to particle number conservation in a closed system. Moreover, the remaining eigenvalues corresponding to the length scale and its temporal derivative are complex conjugated to each other, i.e.~$\lambda_1 = \lambda_2^*$. Comparing Fig.~\ref{fig:vortex_stability}(a) and (b), we see that these eigenvalues can induce an instability into the system; for increasing loss rate the respective real parts become positive and therefore constitute the unstable contribution to the dynamics.
	To investigate this instability in more detail we consider the unstable eigenvalue $\lambda_1$ and determine the corresponding stability boundary. As shown in Fig.~\ref{fig:vortex_stability}(c), in the case of fixed interaction strength the real-part of $\lambda_1$ is always negative for a sufficiently small loss parameter, whereas for larger loss rates the stability boundary can be reached by tuning the pumping strength. Additionally, we observe that for further increasing losses the solution is only stable in the case of a small enough pumping strength. Therefore, we recognize that changing the pumping also leads to a necessary loss change in order to stabilize the solution. The corresponding stability curve can be determined directly from the characteristic polynomial (\ref{eq:characteristic_polynomial}) using the Routh-Hurwitz stability criterion
	\begin{equation}
		\label{eq:stability_criterion_vortex}
		4 - \frac{\gamma^2}{4\hbar^2\omega^2} > \frac{\gamma}{\hbar\omega}\frac{g}{\Gamma}\frac{\ell_\omega^2}{q_0^2} + \frac{4\ell_\omega^4}{q_0^4}\,.
	\end{equation}
	Therefore, we know that, if the stability criterion is fulfilled, the system is stable. At this point, we  emphasize that the stability condition~\eqref{eq:stability_criterion_vortex} and the validity condition $\gamma\leq\hbar\omega$ introduced in Sec.~\ref{sec:SS_vortex} are of different nature. The validity range was introduced heuristically to define a regime in which the Gaussian ansatz~\eqref{eq:ansatz} still represents a good approximation of the density profile. In contrast to that, the stability criterion \eqref{eq:stability_criterion_vortex} is a property of the linearized POEs and does not refer to the quality of the ansatz. Consequently, both conditions are not automatically comparable. However, we note that the instability sets in well inside the validity range.\\
	We now consider how both the characteristic polynomial~\eqref{eq:characteristic_polynomial} and the stability criterion~\eqref{eq:stability_criterion_vortex} are influenced by
	the interaction strength, which can be experimentally tuned independently of the gain and loss of photons by changing the cavity geometry~\cite{Busley_Science2022}. 
	As predicted by Eq.~\eqref{eq:stability_criterion_vortex}, the stable region is increased by increasing the interaction strength, as also shown in the inset of Fig.~\ref{fig:vortex_stability}(d). Additionally, comparing this behavior with the vortex-free case we also see that we get nearly the same stability boundary as follows from Eq.~\eqref{eq:stability_criterion_no_vortex}. Therefore, we conclude that the instability is not induced by the vortex, but is inherited from an instability of the condensate. Moreover, as shown in the inset of Fig.~\ref{fig:vortex_stability}(d), considering instead of a loss change a change in the loss-to-interaction ratio, as also suggested by the stability conditions~\eqref{eq:stability_criterion_vortex} and~\eqref{eq:stability_criterion_no_vortex}, all stability boundary curves collapse to a single one. This indicates that the effective independent quantities defining the stable region of the system are given by the pumping strength and the loss-to-interaction ratio.\\
	Additionally, the real part of the eigenvalue $\lambda_1$ also defines a time scale $\tau = 1/|\text{Re}(\lambda_1)|$, which turns out to be of the order of a few nanoseconds. Comparing this to the condensate lifetime of about $500\,\text{ns}$~\cite{Klaers_Nature2010} shows that in the stable region the vortex reaches its stationary state faster than the condensate decays, so that it is fully established over almost the entire condensate lifetime. Moreover, in the unstable regime the same magnitude of the time scale shows that the vortex solution breaks down within the condensate lifetime. Nevertheless, this behavior distinguishes closed from open-dissipative systems and gives an additional tuning knob that can be possibly used in an experimental set-up as a stabilization mechanism.

	\section{Summary and Outlook}\label{sec:summary}
	Our study of a single-vortex in a harmonically-trapped photon Bose--Einstein condensate has shown that a finite vortex-core size is achievable even though the photon-photon interaction strength is negligibly small. Our ansatz determines that the vortex core size is always smaller than the oscillator length and therefore, in principle, experimentally observable. This behavior was not expected from studies without an external potential in both a closed system~\cite{Pethick_Smith_2008} and a driven-dissipative one~\cite{Krauss_PRR2025}. In particular, we showed that the projection optimization method~\cite{Krauss_PRR2025} provides analytical access to the dynamical description of a system with phase singularities such as in the present case of a singly-charged vortex. Additionally, it allows to generalize previous ansätze for the wave function~\cite{Lewenstein_PRL1996,Lewenstein_PRA1997,Chevy_PRL2002,Stein_NJP2019} as no specific form of the condensate phase has to be assumed. With this, we determined the radial velocity profile in the steady state, which provides a physical definition of the vortex core size based on the hydrodynamic description. This was lacking until now~\cite{Cidrim_PRA2018}. Additionally, we showed that the open-dissipative parameters of pumping and losses can provide an instability of the system. On the other hand, we showed that tuning the open-dissipative parameters can be used to stabilize the system. However, it is still unknown whether the  instability found for larger losses is caused by particle currents that deform the density distribution. Therefore, our results pave the way for further analytical studies of photon condensates with and without vortices. To this end, generalizing the ansatz~\eqref{eq:ansatz} including a possible deformation of the density profile plays a crucial role for future studies.

	\section*{Acknowledgments}
	We thank Fl{\'a}via Braga Ramos, Nikolai Kaschewski, and Leon Mixa for fruitful discussions. Furthermore, we acknowledge financial support by the Deutsche Forschungsgemeinschaft (DFG, German Research Foundation) via the Collaborative Research Center SFB/TR185 (Project No. 277625399). And this work was supported by CNPq (Conselho Nacional de Desenvolvimento Científico e Tecnológico, National Council for Scientific and Technological Development), Brazil and DAAD-CAPES PROBRAL, Brazil Grant No. 88887.627948/2021-00. T. Bates acknowledges the support of a scholarship from the German Academic Exchange Service (DAAD). Finally, F. E. A. dos Santos thanks CNPq for support through Bolsa de produtividade em Pesquisa (Research productivity scholarship) Grant No. 310161/2023-1.

	\appendix
	\numberwithin{equation}{section}
	\section{Vortex-free Projection Optimization Solution}\label{app:no_vortex}
	In this section, we consider a harmonically trapped photon BEC without any phase singularities. The aim of this section is to study both the stationary solution as well as its linear stability. To this end, we derive in App.~\ref{app:POE_no_vortex} the dynamical projection optimization equations. App.~\ref{app:no_vortex_ss} then determines the stationary solution, where we briefly compare this to the case of a single vortex determined in Sec.~\ref{sec:SS_vortex}. Lastly, we determine in App.~\ref{app:linear_stability_no_vortex} the linear stability of the steady-state solution, which allows us to determine the breathing mode of the open-dissipative photon condensate.
	\subsection{Projection Optimization Equations}\label{app:POE_no_vortex}
	Here, we derive equations that approximately solve the cGPE~\eqref{eq:cGPE} without any singularities. To this end, we use the projection optimization method introduced in Sec.~\ref{sec:method} with the ansatz~\eqref{eq:ansatz} for $\ell=0$. This ansatz allows to determine the radial phase without any assumption and therefore generalizes well-established findings for both closed~\cite{Lewenstein_PRL1996,Lewenstein_PRA1997} and open-dissipative systems~\cite{Stein_NJP2019}. Also note that in this case the length scale $q$ determines the size of the condensate and is known as the width of the condensate. Therefore, it is a distinct length scale compared to the single-vortex case.\\
	Following the steps of Sec.~\ref{sec:POE_vortex} yields at first the radial velocity and phase profiles, given by
	\begin{align}
		\label{eq:vel_no_vortex}
		v_{\rm R} &= \frac{r\partial_t q}{q} + \left(
		\frac{\gamma}{\hbar}  - \frac{\partial_t N}{N}\right)
		\frac{q^2}{2r}
		\left(e^{-\frac{r^2}{q^2}} - 1\right)\,,\\
		\label{eq:phase_no_vortex}
		\varphi_{\rm R} &= 
		\frac{m}{\hbar} \left\{ \frac{r^2\partial_t q}{2q} - \left(\frac{\gamma}{\hbar}-\frac{\partial_t N}{N}\right)
		\frac{q^2}{4}
		\left[\gamma_{\rm EM} + E_1\left(\frac{r^2}{q^2}\right) + \ln\left(\frac{r^2}{q^2}\right)\right]\right\}\,.
	\end{align}
	At this point, we recognize that Eqs.~\eqref{eq:vel_no_vortex} and~\eqref{eq:phase_no_vortex} reduce in the closed system limit to the well-known results from Refs.~\cite{Lewenstein_PRL1996,Lewenstein_PRA1997}. But we have to emphasize that 
	we achieve this result without performing an explicit ansatz for the radial phase. 
	Within the projection optimization method the radial phase follows from the ansatz of the condensate density. In the same spirit we also generalize here the cumulant approach from Ref.~\cite{Stein_NJP2019}.\\
	With this the POEs for the particle number, the condensate width, and the chemical potential read
	\begin{align}
		\label{eq:N_no_vortex}
		\partial_t N &= N\left(\frac{\gamma}{\hbar} - \frac{\Gamma N}{2\pi\hbar q^2}\right)\,,\\
		\label{eq:q_no_vortex}
		\partial^2_t q &= \frac{\hbar^2}{m^2 q^3} - \omega^2 q + \frac{g N}{2\pi m q^3} + \frac{\Gamma \partial_t N}{8 \pi \hbar q} + \frac{[\ln(4/3) - 1 / 3] \Gamma^2N^2}{16\pi^2\hbar^2q^3} \,,\\
		\label{eq:mu_no_vortex}
		\partial_t\mu &=  \frac{m\omega q\partial^2_t q}{2} + \frac{\hbar^2\omega}{2m q^2} + \frac{m\omega ^3 q^2}{2} + \frac{\omega gN}{2\pi q^2} - \frac{\ln (2) m\omega \Gamma \partial_t N}{8\pi\hbar} + \frac{\ln(4/3) m\omega \Gamma^2 N^2}{32\pi^2\hbar^2q^2}\,.
	\end{align}
	Note that in order to simplify the notation we do not distinguish between the vortex case from Sec.~\ref{sec:POE_vortex} and the vortex-free case.\\
	Additionally, we remark that Eqs.~\eqref{eq:N_no_vortex}--\eqref{eq:mu_no_vortex} have the same formal structure as in the vortex case. Therefore, also in the present case the chemical potential is purely determined by the particle number and the condensate width, so the system of coupled equations to be solved is given by Eqs.~\eqref{eq:N_no_vortex} and~\eqref{eq:q_no_vortex}.

	\subsection{Stationary State}\label{app:no_vortex_ss}
	In this section, we derive the steady-state solution of the vortex-free POEs. To this end, we make the ansatz of time-independent particle number and condensate width, i.e.~$N(t)=N_0$, $q(t)=q_0$, as well as a linear ansatz for the time evolution of the chemical potential $\mu(t)=\omega\mu_0 t$. Doing so, the POEs~\eqref{eq:N_no_vortex}--\eqref{eq:mu_no_vortex} reduce to algebraic equations, whose solution is given by
	\begin{align}
		\label{eq:q_no_vortex_ss}
		q^2_{0,\pm} &= \frac{-\frac{g\gamma}{2\hbar\omega\Gamma} \pm \sqrt{\left(\frac{g\gamma}{2\hbar\omega\Gamma}\right)^2 - \frac{\gamma^2}{4\hbar^2\omega^2}\left[\ln\left(\frac{4}{3}\right) - \frac{1}{3} - \frac{4\hbar^2\omega^2}{\gamma^2}\right]}}{\frac{\gamma^2}{4\hbar^2\omega^2}\left[\ln\left(\frac{4}{3}\right) - \frac{1}{3} - \frac{4\hbar^2\omega^2}{\gamma^2}\right]}\ell_\omega^2\,,\\
		\label{eq:N_no_vortex_ss}
		N_0 &= \frac{2\pi\gamma q^2_{0,\pm}}{\Gamma}\,,\\
		\label{eq:mu_no_vortex_ss}
		\mu_0 &= \frac{\hbar^2}{2m q^2_{0,\pm}} + \frac{m\omega^2q^2_{0,\pm}}{2} + \frac{g\gamma}{\Gamma} + \frac{m\gamma^2 q^2_{0,\pm}}{8\hbar^2}\ln\left(\frac{4}{3}\right)\,.
	\end{align}
	\begin{figure}[t!]
		\centering
		\includegraphics[width=\columnwidth]{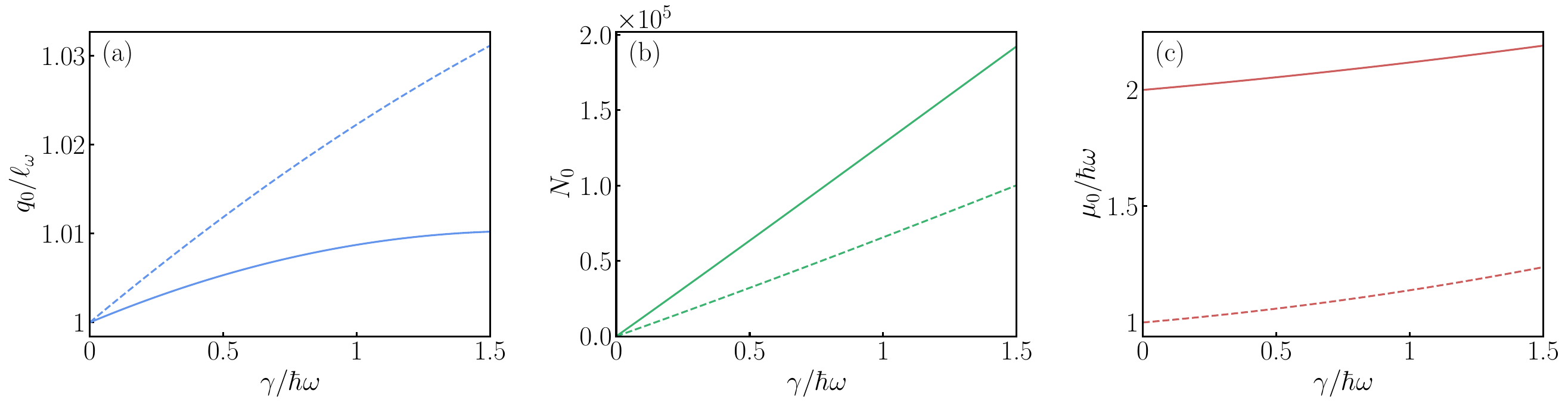}
		\caption{Vortex-free projection optimization parameters (dashed lines) determined in Eqs.~\eqref{eq:q_no_vortex_ss}--\eqref{eq:mu_no_vortex_ss} and single-vortex optimization parameters following from Eqs.~\eqref{eq:q_vortex_ss}--\eqref{eq:mu_vortex_ss} (solid lines). Evaluated for fixed interaction $g = 10^{-5}\cdot \hbar^2/m$ and losses $\Gamma = 10^{-4}\cdot \hbar^2/m$.}
		\label{fig:proj_params_no_vortex}
	\end{figure}
	First, we note that the stationary solution of the vortex-free system differs from the vortex case only in the prefactors, whereas the overall structure remains the same. So we conclude that the structure of the equations is given by the chosen ansatz~\eqref{eq:ansatz}. A comparison of both cases is shown in Fig.~\ref{fig:proj_params_no_vortex}. We note that the overall system size, determined by the condensate width, increases much faster than its vortex counterpart. In contrast to that the particle number as well as the chemical potential remain below the single-vortex case. The particle number grows linearly with the pumping strength, coinciding with the expectation from the rate-equations~\cite{Krauss_EPJST2026}. We also note that the chemical potential fulfills Eq.~\eqref{eq:chem_pot_expand}.\\
	We conclude the stationary-state analysis with the radial velocity and phase profiles. Both follow from the respective dynamic Eqs.~\eqref{eq:vel_no_vortex},~\eqref{eq:phase_no_vortex} and read
	\begin{align}
		\label{eq:vel_no_vortex_ss}
		v_{\rm R,0} &= \frac{\gamma q^2}{2 \hbar r} \left(e^{-\frac{r^2}{q^2}} - 1\right)\,,\\
		\label{eq:phase_no_vortex_ss}
		\varphi_{\rm R,0} &= -\frac{m\gamma q^2}{4\hbar^2}\left[\gamma_{\rm EM} + E_1\left(\frac{r^2}{q^2}\right) + \ln\left(\frac{r^2}{q^2}\right)\right]\,.
	\end{align}
	We immediately notice that, in contrast to the vortex, the radial phase in the vortex-free case is always negative. Consequently, this also holds for the radial velocity profile. Apart from the overall height, which is given by the pumping strength as evident from Eqs.~\eqref{eq:vel_no_vortex_ss} and~\eqref{eq:phase_no_vortex_ss}, the shape is depicted in Fig.~\ref{fig:vortex_anatomy}(a) and (b). Moreover, for the physical discussion of the radial velocity and phase profiles we refer to Sec.~\ref{sec:SS_vortex}.

	\subsection{Linear Stability and Breathing Mode}\label{app:linear_stability_no_vortex}
\begin{figure}[t!]
		\centering
		\includegraphics[width=\columnwidth]{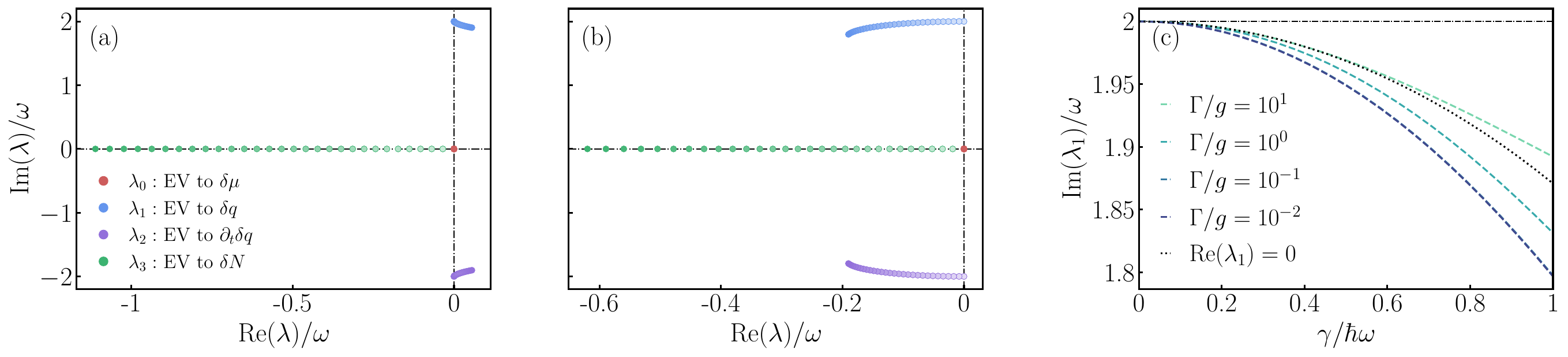}
		\caption{Panel (a) and (b) show eigenvalues of matrix $B$ as function of pumping strength. Bright colors denote small pumping with smallest value $\gamma/\hbar\omega = 0$, whereas strong colors denote large pumping with  maximum being $\gamma/\hbar\omega = 1$. In both Panels we have $g = 10^{-5}\cdot \hbar^2/m$, whereas in the left (right) Panel $\Gamma = 10^{-3}\cdot \hbar^2/m$ ($\Gamma = 10^{-7}\cdot \hbar^2/m$) is used. Panel (c) depicts imaginary part of eigenvalue $\lambda_1$ for different loss-to-interaction rates. The dotted line shows Eq.~\eqref{eq:breathing_freq} at vanishing ${\rm Re}(\lambda_1)$.}
		\label{fig:stability_no_vortex}
	\end{figure}

	In this section, we determine the stability equations that define whether the solution found in App.~\ref{app:no_vortex_ss} is stable or not. Additionally, this allows us to determine the breathing mode frequency of the condensate, which stems from the oscillation of the condensate width. To this end, we consider small perturbations $\delta\bullet$ from the stationary state, i.e.~we make the ansatz $N(t) = N_0 + \delta N(t)$, $q(t) = q_0 + \delta q(t)$, and $\mu(t) = \omega\mu_0 t + \delta\mu(t)$ with the perturbations. Expanding the POEs~\eqref{eq:N_no_vortex}--\eqref{eq:mu_no_vortex} up to linear order in the perturbations, yields the first-order differential equation $\partial_t \boldsymbol{y} = B \boldsymbol{y}$ with the vector $\boldsymbol{y}^{\rm T} = \left(\delta N(t), \delta q(t), \partial_t \delta q(t), \delta \mu(t)\right)$ and the matrix $B\in {\rm Mat}(4\times 4, \mathbb{R})$. This has the formal solution
	\begin{equation}
		\label{eq:stability_solution_no_vortex}
		\boldsymbol{y}(t) = e^{B t}\boldsymbol{y}_0\,,
	\end{equation}
	with the initial value $\boldsymbol{y}_0$. Whether the formal solution~\eqref{eq:stability_solution_no_vortex} remains bounded is determined by the eigenvalues of the matrix $B$. Analogously to Sec.~\ref{sec:Stability_vortex} we determine the eigenvalues via the roots of the 
	corresponding
	characteristic polynomial
	\begin{equation}
		\label{eq:characteristic_polynomial_no_vortex}
		\chi_B(\lambda) = \omega^4 \frac{\lambda}{\omega}\left[\left(\frac{\lambda}{\omega}\right)^3 + \frac{\gamma}{\hbar\omega}\left(\frac{\lambda}{\omega}\right)^2 + \left(4-\frac{\gamma^2}{2\hbar^2\omega^2}\right)\frac{\lambda}{\omega} + \frac{\gamma^2 g \ell_\omega^2}{\hbar^2\omega^2\Gamma q_0^2}\left(2 + \frac{4\hbar\omega\Gamma\ell_\omega^2}{g\gamma q_0^2} \right)\right]\,.
	\end{equation}
	Analogously to the vortex case, the eigenvalue corresponding to the perturbation of the chemical potential always vanishes, i.e.~$\lambda_0 = 0$ stemming from the $U(1)$ symmetry of the cGPE~\eqref{eq:cGPE}. The remaining eigenvalues are visualized in Fig.~\ref{fig:stability_no_vortex}(a) and (b). We see that the eigenvalue for the perturbation of the particle number remains always real-valued and negative, whereas in the closed system limit the eigenvalue becomes zero due to particle number conservation. However, the eigenvalues corresponding to the condensate width and its temporal derivative are complex conjugate to each other, i.e.~$\lambda_1 = \lambda_2^*$, and get for larger loss rates a positive real part. Therefore, the perturbation of the condensate width is the unstable contribution to the vortex-free dynamics, where the instability boundary follows from the Routh-Hurwitz criterion given by
	\begin{equation}
		\label{eq:stability_criterion_no_vortex}
		4 - \frac{\gamma^2}{2\hbar^2\omega^2} > \frac{2\gamma}{\hbar\omega}\frac{g}{\Gamma}\frac{\ell_\omega^2}{q_0^2} + \frac{4\ell_\omega^4}{q_0^4}\,.
	\end{equation}
	A detailed discussion of the instability is presented in Sec.~\ref{sec:Stability_vortex}.\\
	Moreover, the imaginary part of these eigenvalues corresponds to the frequency of the breathing mode. From the discussion above we know that the three non-vanishing roots of Eq.~\eqref{eq:characteristic_polynomial_no_vortex} are given by one real root and a complex conjugate pair. Therefore, rewriting Eq.~\eqref{eq:characteristic_polynomial_no_vortex} using this information allows to determine the breathing frequency as 
	\begin{equation}
		\label{eq:breathing_exact}
		\left[\frac{{\rm Im}(\lambda_1)}{\omega}\right]^2 - \left(4 - \frac{\gamma^2}{2\hbar^2\omega^2}\right) = \frac{{\rm Re}(\lambda_1)}{\omega}\left(\frac{2\gamma}{\hbar\omega} + 3\frac{{\rm Re}(\lambda_1)}{\omega}\right)\,.
	\end{equation}
	Note that Eq.~\eqref{eq:breathing_exact} depends on interaction and losses only implicitly through the real part ${\rm Re}(\lambda_1)$. As the latter is a function of both the pumping and the interaction-to-loss ratio, the curves do not collapse onto each other for different parameters. This is in contrast to the stability boundary shown in the inset of Fig.~\ref{fig:vortex_stability}(d). However, at the stability boundary the breathing frequency is independent of the loss-to-interaction ratio and given by
	\begin{equation}
		\label{eq:breathing_freq}
		\left[\frac{{\rm Im}(\lambda_1)}{\omega}\right] = \sqrt{4 - \frac{\gamma^2}{2\hbar^2\omega^2}} \,.
	\end{equation}
	The corresponding behavior is depicted in Fig.~\ref{fig:stability_no_vortex}(c). It is shown that the frequency decreases with increasing pumping power, agreeing with the literature~\cite{Stein_NJP2019}. For vanishing pumping the breathing frequency approaches the value $2\omega$ known from closed systems~\cite{Chevy_PRL2002,Lewenstein_PRL1996,Lewenstein_PRA1997}. Furthermore, we also see the implicit dependence of the breathing frequency on the loss-to-interaction ratio. Tuning this ratio shows that a decrease yields also a decrease of the frequency for increasing pumping strength. Since Eq.~\eqref{eq:breathing_freq} corresponds to a vanishing real part of the eigenvalue $\lambda_1$, it also indicates the onset of instability. Frequencies above this threshold correspond to unstable steady states and those below remain stable. Consequently, lowering the loss-to-interaction ratio decreases the breathing frequency and stabilizes the condensate simultaneously.

	\section{Hydrodynamic Description}\label{app:hydrodynamics}
	Rather than solving the time-dependent cGPE approximately, as was done above, further insight into the condensate dynamics is gained from its hydrodynamic formulation~\cite{Lamb_1975,Lifschitz_1987}. This approach is based on the Madelung representation of the condensate wave function, given by
	\begin{equation}
		\label{eq:madelung}
		\Psi({\bf r},t) = \sqrt{n({\bf r},t)} e^{i\Phi({\bf r},t)}\,,
	\end{equation}
	where $n({\bf r},t)$ denotes the condensate density and $\Phi({\bf r},t)$ its phase.\\
	Inserting Eq.~\eqref{eq:madelung} into the cGPE~\eqref{eq:cGPE}, separating real and imaginary parts, and using $\boldsymbol{v}({\bf r},t) = \hbar\boldsymbol{\nabla}\Phi({\bf r},t)/m$~\cite{Pethick_Smith_2008}, yields the following two equations
	\begin{align}
		\label{eq:continuity_eq}
		\frac{\partial n}{\partial t} &= n(\gamma - \Gamma n) - \boldsymbol{\nabla}\cdot(n\boldsymbol{v})\,,\\
		\label{eq:Newton_eq}
		m\frac{\partial \boldsymbol{v}}{\partial t} &= \boldsymbol{\nabla} \left[\frac{\hbar^2}{2m}\frac{\boldsymbol{\nabla}^2\sqrt{n}}{\sqrt{n}} - \frac{1}{2} m \boldsymbol{v}^2 - \frac{1}{2} m \omega^2 {\bf r}^2 - gn\right]\,.
	\end{align}
	Here, Eq.~\eqref{eq:continuity_eq} is a generalized continuity equation whose first term on the right-hand side accounts for particle gain and loss, while Eq.~\eqref{eq:Newton_eq} is an Euler equation~\cite{Lifschitz_1987}. The velocity field thus obeys the dynamics of an ideal flow driven by the gradients of the kinetic, trapping and interaction energies, together with the quantum pressure associated with the Bohm potential
	\begin{equation}
		\label{eq:quantum_pressure}
		V_{\rm qp} = -\frac{\hbar^2}{2m}\frac{\boldsymbol{\nabla}^2\sqrt{n}}{\sqrt{n}}\,.
	\end{equation}
	Rewriting Eq.~\eqref{eq:Newton_eq} yields in the absence of any vorticity the quantum Bernoulli equation
	\begin{equation}
		\label{eq:Bernoulli_eq}
		\frac{\partial \boldsymbol{v}}{\partial t} + \left(\boldsymbol{v}\cdot\boldsymbol{\nabla}\right)\boldsymbol{v} = \boldsymbol{\nabla} \left[\frac{\hbar^2}{2m^2}\frac{\boldsymbol{\nabla}^2\sqrt{n}}{\sqrt{n}} - \frac{1}{2} \omega^2 {\bf r}^2 - \frac{gn}{m}\right] %+ \left(\boldsymbol{\nabla}\times\boldsymbol{v}\right)\times\boldsymbol{v}
		\,.
	\end{equation}
	It exposes the three pressure contributions given by the mechanical pressure of the external trap, the interaction pressure due to repulsive particle--particle interactions, and the quantum pressure, which measures the resistance of the fluid against density deformation. Note that in the presence of vorticity an additional term appears in Eq~\eqref{eq:Bernoulli_eq} due to the multivalued nature of the phase, see for instance Ref.~\cite{Ednilson_PRA2016}.
	\begin{figure}[t!]
		\centering
		\includegraphics[width=\columnwidth]{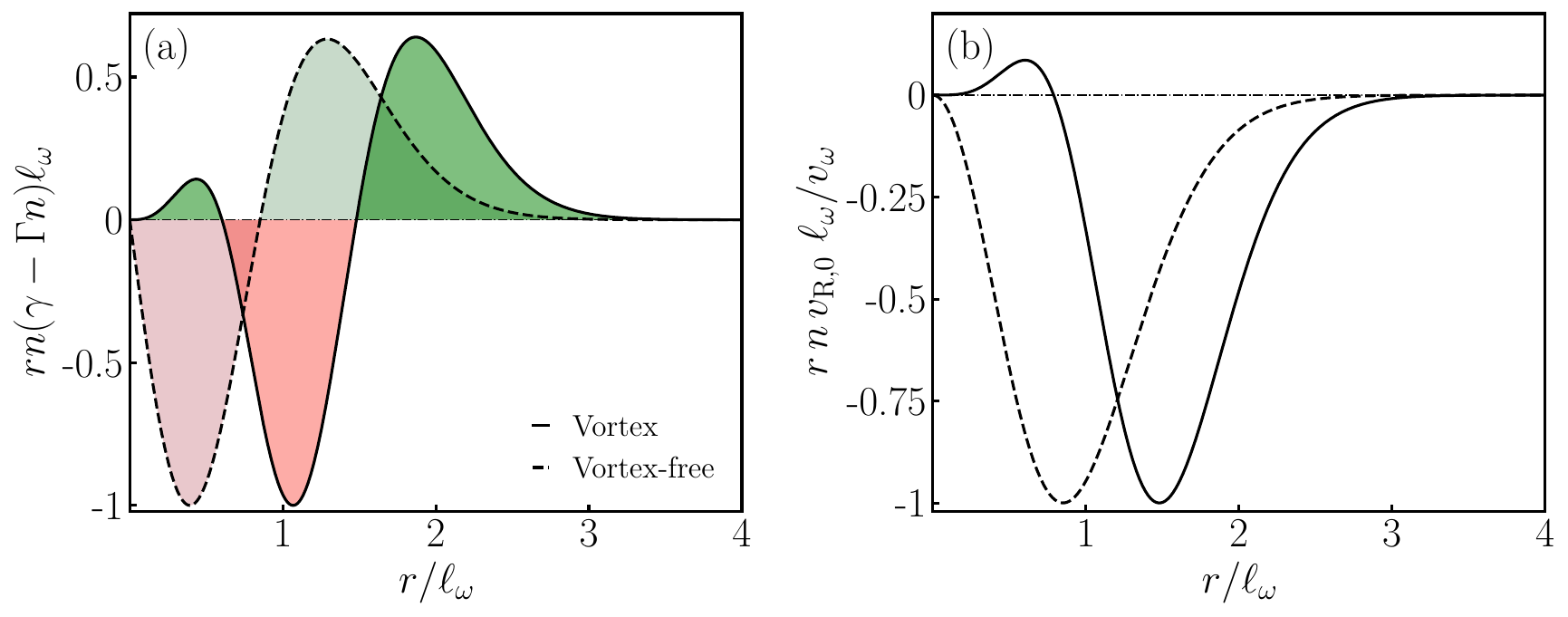}
		\caption{Panel (a) shows local net source, i.e.~number of photons created per unit time and unit radius, for vortex (solid) and vortex-free (dashed) steady state determined from Eq.~\eqref{eq:ansatz} with respective $\ell$ and projection optimization parameters according to Sec.~\ref{sec:POE_vortex} (vortex solution) and App.~\ref{app:no_vortex_ss} (vortex-free). Shading marks the regions of net gain (green) and net loss (red) of the vortex and vortex-free solutions. Panel (b) shows corresponding radial integral, according to Eq.~\eqref{eq:continuity_eq_ss}. In both Panels curves are normalized to their own maximum value. For every curve $\gamma/\hbar\omega = 1.0$, $\Gamma m/\hbar^2 = 10^{-4}$, and $gm/\hbar^2 = 10^{-5}$ is used.}
		\label{fig:source_sink}
	\end{figure}

	\subsection{Mass Balance and Momentum Balance}\label{app:stationary_hydrodynamics}
	For a stationary condensate, which is described by $\Psi({\bf r},t) = \sqrt{n({\bf r})} e^{i\left[\Phi({\bf r})-\mu t/\hbar\right]}$, Eqs.~\eqref{eq:continuity_eq} and~\eqref{eq:Newton_eq} act differently on the condensate density and the velocity field. Equation~\eqref{eq:Newton_eq} fixes the shape of the density profile, whereas the velocity field is determined from Eq.~\eqref{eq:continuity_eq}. For a radially symmetric steady state the velocity field is determined by
	\begin{equation}
		\label{eq:continuity_eq_ss}
		r\,n\,v_{\rm R,0} = \int_0^{r}\,{\rm d}r^\prime\,r^\prime n(r^\prime)\Big[\gamma - \Gamma n(r^\prime)\Big]\,.
	\end{equation}
	The left-hand side of Eq.~\eqref{eq:continuity_eq_ss} describes the total number of particles created per time unit inside a disc of radius $r$. If it vanishes at a distance $r_0$, consequently, the integrated net gain has to balance the particle losses. This leads to 
	\begin{equation}
		\label{eq:stagnation_condition}
		\int_0^{r_0}\,{\rm d}r^\prime\,r^\prime n(r^\prime)\left[\gamma - \Gamma n(r^\prime)\right] = 0\,,
	\end{equation}
	which is the condition of a vanishing radial velocity in order to fulfill Eq.~\eqref{eq:continuity_eq_ss}. This condition is caused by the balance of mass in the system and fixes the zero points of the radial velocity, and has to be distinguished from the point of maximum density $r_{\rm peak}$, defined via $\partial_r n |_{r_{\rm peak}} = 0$, which is caused by the balance of momenta and set by Eq.~\eqref{eq:Newton_eq}. Both conditions, therefore, originate from the different equations describing mass and momentum balance, respectively.\\
	The mechanism behind Eq.~\eqref{eq:stagnation_condition} is illustrated in Fig.~\ref{fig:source_sink}. Since gain and loss balance locally at the density $\gamma/\Gamma$, the condensate separates into a gain-dominated core, a loss-dominated ring, and a gain-dominated dilute tail, see Fig.~\ref{fig:source_sink}(a). Photons created in the depleted core flow outwards and photons created in the tail flow inwards, both being absorbed in the ring, so that the integrated flux shown in Fig.~\ref{fig:source_sink}(b) changes sign. That distance lies inside the ring of accumulated particles, which is a consequence of the small vortex-core area, whose outward flux is compensated over only a thin inner part of the loss-dominated ring. In the vortex-free case the peak density exceeds $\gamma/\Gamma$ already at the trap center, the innermost region is a sink rather than a source, and the integrated flux never changes sign.

	$\mbox{}$\\
	\printbibliography

\end{document}